# Sustainable business models: a review

Saeed Nosratabadi [1], Amir Mosavi [2,3,7], Shahaboddin Shamshirband [4,5]*, Edmundas Kazimieras Zavadskas [6], Andry Rakotonirainy [7]. Kwok Wing Chau[8]

[1] Institute of Business Studies, Szent Istvan University, Godollo, Hungary; Saeed.Nosratabadi@phd.uni-szie.hu
[2] Institute of Automation, Kando Kalman Faculty of Electrical Engineering, Obuda University, 1034 Budapest, Hungary; amir.mosavi@kvk.uni-obuda.hu
[3] School of the Built Environment, Oxford Brookes University, Oxford, UK;
[4] Faculty of Information Technology, Ton Duc Thang University, Ho Chi Minh City, Viet Nam;
[5] Department for Management of Science and Technology Development, Ton Duc Thang University, Ho Chi Minh City, Vietnam;
[6] Institute of Sustainable Construction, Vilnius Gediminas Technical University, LT-10223 Vilnius, Lithuania; edmundas.zavadskas@vgtu.lt
[7] The Queensland University of Technology, Institute of Health and Biomedical Innovation, 60 Musk Avenue, Kelvin Grove, Queensland 4059, Australia; r.andry@qut.edu.au
[8] Department of Civil and Environmental Engineering, Hong Kong Polytechnic University, Hung Hom, Hong Kong
* Correspondence: shahaboddin.shamshirband@tdtu.edu.vn

**Abstract:** During the past two decades of the e-commerce growth the concept of business model has become increasingly popular. More recently, the research on this realm has grown rapidly with a diverse research activity covering a wide range of application areas. Considering the sustainable development goals the innovative business models have brought a competitive advantage to improve the sustainability performance of organizations. The concept of the sustainable business model describes the rationale of how an organization creates, delivers, and captures value, in economic, social, cultural or other contexts in a sustainable way. The process of sustainable business model construction forms an innovative part of business strategy. Different industries and businesses have utilized sustainable business models' concept to satisfy their economic, environmental and social goals simultaneously. However, the success, popularity, and the progress of sustainable business models in different application domains are not clear. To explore this issue, this research provides a comprehensive review of sustainable business models literature in various application areas. Notable sustainable business models are identified and further classified in fourteen unique categories, and in every category, the progress -either failure or success- has been reviewed and the research gaps are discussed. Taxonomy of the applications includes innovation, management and marketing, entrepreneurship, energy, fashion, healthcare, agri-food, supply chain management, circular economy, developing countries, engineering, construction and real estate, mobility and transportation, and hospitality. The key contribution of this study is to provide an insight into the state of the art of sustainable business models in various application areas and future research directions. This paper concludes that popularity and the success rate of sustainable business models in all application domains have been increased along with the increasing use of advanced technologies.

## 1. Introduction

The business model concept is an abstract representation of the value flow and the interactions between value elements of an organizational unit. The essential value elements of organizations are concerned with proposition, creation, delivering, and capturing value. A simplified way of communicating the connection and function of these elements is vital in the success of any business [1]. For this purpose, the concept of business model has originated to facilitate explanation of complex business ideas more efficiently. Through a business model, the business workflow is communicated to investors in detail within a short time frame [2]. In fact, the effective representation of planning, analysis, communication, and implementation of organizational complex units' performance are reported as one of the major reasons behind the popularity of business models [3]. Geissdoerfer et al. [4] present a detailed review of the different types and various definitions of business models where a vast number of definitions are presented. Model of an organizational system [5], a simplified characteristic of business concept [6], and a reduced scope of business [7] are suggested as the various types of business models. For decades the vital sustainability issues with their major societal and environmental effects influencing human beings and nature had not been the priorities of most business model types. Nevertheless, business models, for achieving the sustainability goals of companies have finally become under pressure to transform into a more sustainable economic system.

Internationalization along with the urge to keep up with the sustainable development goals has made the worldwide competition among the firms more complex with conventional business models struggling with finding appropriate solutions. In this context, the alternative concept of sustainable business model has brought competitive advantage to organizations through empowering the conventional business models meeting the sustainable development goals while maintaining productivity and profitability [8, 9]. Thus, creating value for the triple bottom line, i.e. economic, society and environment, has been the ultimate goal of sustainable business models [10]. Sustainable business models have the great potential to incorporate the principles of sustainability and integrate sustainability goals into the value proposition, value creation, and value capture activities of businesses [11]. Sustainable business models aim at employing proactive multi-stakeholder management, innovation, and long-term perspective to meet sustainability goals. Sustainable business models, therefore, have been effectively contributing in reducing the harmful effects of business activities on the environment and society through providing solutions to help firms meet their economic and sustainability goals simultaneously [12]. Thus, the concept of sustainable business model has emerged to provide a platform for integrating sustainability considerations [13]. From this perspective circular business models [14] share similarity with sustainable business models. However, they include additional characteristics which are mainly concerned with slowing, intensifying, and narrowing resource loops [4].

The review paper of Evans et al. [15] shows how sustainable business models have helped businesses to achieve their sustainability ambitions. Further research, e.g. Boons et al. [11], Geissdoerfer et al. [4] and Schaltegger et al. [9] provide a collection of the definitions to the concept of sustainable business model. According to Lüdeke-Freund [16], sustainable business models are tools for delivering social and environmental sustainability to the industrial systems. Whilst, there are constraints for understanding the sustainable business models and the available innovative alternatives for transformation to sustainability [14]. Despite numerous research on sustainable business models in the literature, there is no comprehensive picture of how firms in different industries can implement sustainability in their business models. Although there exists literature on the definitions and overview of the concept of sustainable business model, there is a research gap in the progress and evaluation of the performance of sustainable business models in each specific application domain. The spread and effectiveness of sustainable business models in business domains have not been identified. Furthermore, the applicability, popularity, success, and future trends in various business domains have been not discussed yet. Consequently, the contribution of this article is to present a classification of the widespread applications of sustainable business models in addition to an in-depth investigation of various application domains considering the success and failure cases.

The rest of this paper is structured as follow. Section two presents the methodology of the review. Section three presents the taxonomy of the research and the review's initial report. Further, in the fourteen subsections, the applications of sustainable business models in the individual categories are presented. Section four and five respectively present the discussion and conclusions of the research.

## 2. Methodology

The primary goal of this literature survey is to present the state of the art of sustainable business models in the individual application areas. Accordingly, the research methodology has been developed to identify, classify and review the notable peer-reviewed articles in the design and implementation of sustainable business models in top-level subject fields. Using the Thomson Reuters Web-of-Science (WoS) and Elsevier Scopus for implementation of the search queries would assure that any paper in the database would meet four criteria of quality measure, i.e., source normalized impact per paper (SNIP), CiteScore, SCImago journal rank (SJR), and h-index. Through the search query of "business model*" and "sustainab*" for title, abstract and keywords the relevant literature are identified. The query of (TITLE-ABS-KEY ("business model*") AND TITLE-ABS-KEY (sustainab*)) would result in 6,330 document results (3,494 document in the Scopus and 2,836 documents in WoS). However, through auxiliary search keywords such as "sustainable development" in all fields of the paper, we make sure that the most relevant papers are identified, and the paper significantly contributes to the definition of sustainable development. Consequently, the alternative search query of (TITLE-ABS-KEY ("business model*") AND TITLE-ABS-KEY (sustainab*) AND ALL ("sustainable development")) would result in 1,584 document results (875 results in the Scopus and 709 results in the WoS) which form our initial database. Reading in detail the articles' relevancy downed the numbers to 66 articles for final consideration. The research methodology follows a comprehensive and structured workflow based on a systematic database search and cross-reference snowballing. The flowchart of the research methodology is presented in figure 1. The method is considered as a modified version of review proposed by [17].

In the first step, the search queries explore the Thomson Reuters Web-of-Science and Elsevier Scopus databases. In the second step, the abstract and keywords of the identified articles are browsed to identify the relevant literature and exclude the irrelevant ones. In step three the database of the relevant articles is created. In step four, the article is carefully read, and the category of the application is identified accordingly. In this step, the expert-based knowledge and the initial preferences would influence the number and the type of the categories. In step five we decide on generating a new category and export the article in a new table of application domain or pass the article to step six where a category would host an article in its table. Once a category is created for a new article, in step seven, we pass that article to that category. In step eight we save the content of our database in various categories, update the content of the tables, and review the papers. This workflow will be repeated until sorting out all the papers.

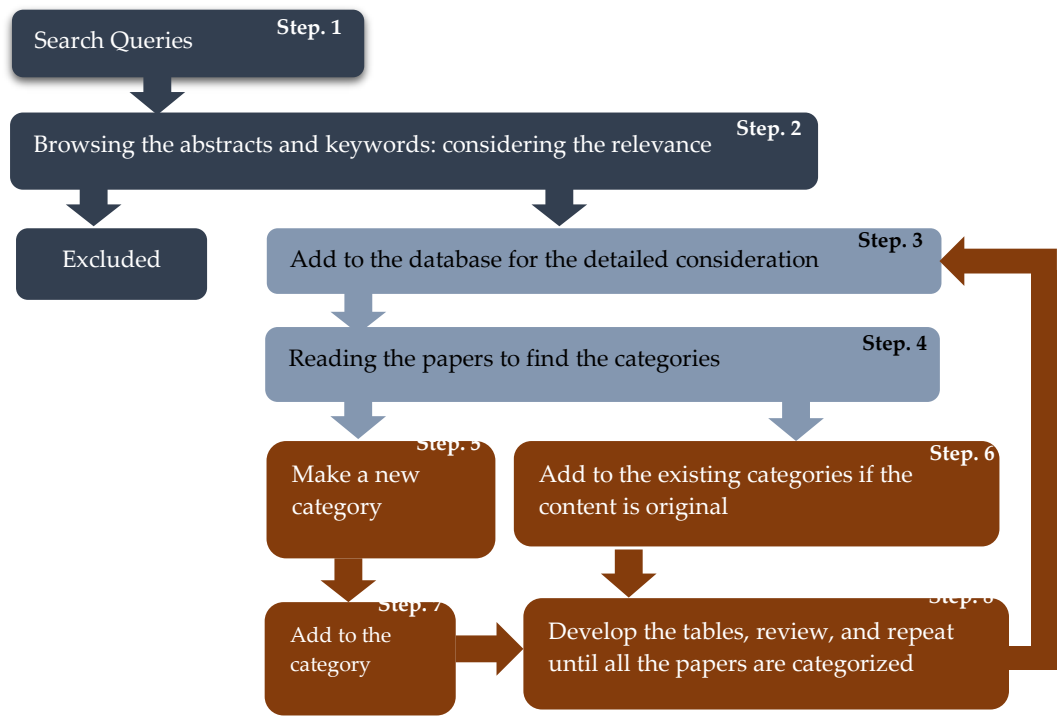

**Figure 1.** Flowchart of the methodology of research

## 3. Review

Studying the initial database communicates interesting information. As it is depicted in figure 2, the number of publications in a sustainable business model is remarkably increased during the past two decades. Where only 2 documents published in this area in 2002 in compare with 62 documents in 2016 and 74 documents in 2017. It implies that one of the major solutions for sustainable development is a sustainable business model and the firms have utilized sustainable business models to have eco-socio friendly business activities.

Furthermore, the nature of a particular business is very determinant in the approaches that the firms can select for their sustainable business models. In fact, the solutions provided in the literature present different characteristics for the implementation of sustainable business models according to the business domain. On the other hand, that implementation of a sustainable business model implies new challenges, innovation or adjustment with new activities. Since sustainability deals with triple bottom line factors, in addition to the financial benefit [14, 18, 19] the benefits of multiple-stakeholders such as customers, suppliers, shareholders etc. have been considered in sustainable development. Therefore, the transition toward sustainable business models requires to look beyond the entity of the firm and it needs innovation activities to create value for the triple bottom line. Hence, incremental changes are insufficient to address sustainable development challenges [20, 21]. The current study provides insights on the research path of the sustainable business model. The paper, as a literature review, increases the knowledge of how different industries, sectors, research areas apply sustainable business models in order to achieve sustainability goals and progress towards sustainable development.

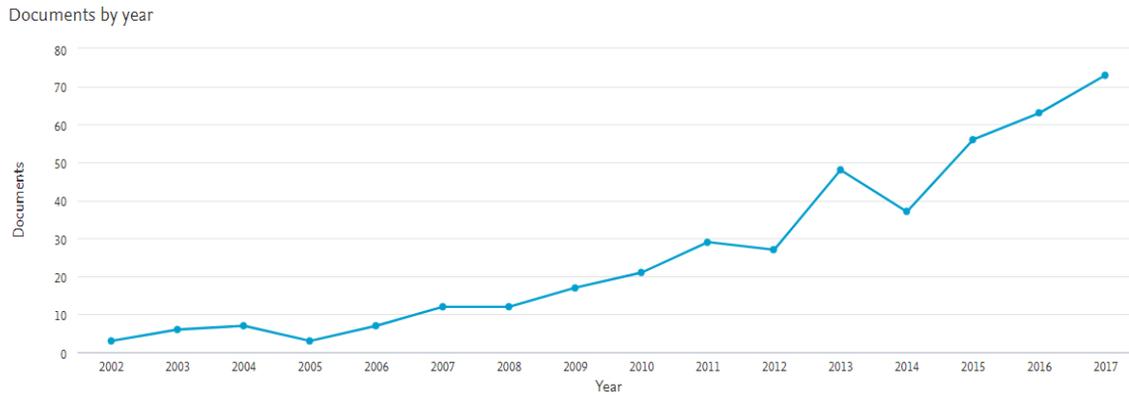

**Figure 2**. The number of publications on sustainable business models from 1999-2018.

The initial database of the literature disclosed that the number of publications on sustainable business models, as it is also shown in figure 3, has been increasing year by year. It is diagnosed that the journal of Sustainability (with 44 documents), Journal of Cleaner Production (with 49 documents), Procedia CIRP (with 25 documents), Water Resources Management (with 15 documents), Environment Development and Sustainability (with 24 documents) are the major journals that have published the results and findings of research on sustainable business models. Figure 3 also clarifies that the number of documents published in these journals is increasing particularly from 2014 onward. Journals of Cleaner Production and Sustainability have had the most share of these trend as they have published the most publications in sustainable business models.

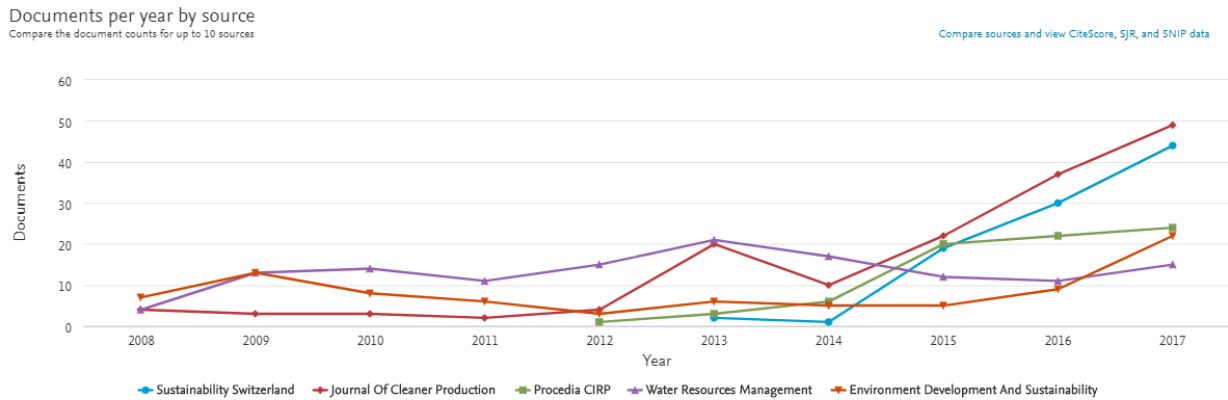

**Figure 3.** The number of publications in the sustainability business model in different journals.

In figure 4, the data related to the different subject areas have utilized sustainable business models in their either title, or abstract, or keywords. The pie chart on the left side refers to the documents published from 2007 to 2018 and the right-side pie chart refers to the documents published from 2015 onward. According to figure 3, 'Business, Management and Accounting (with 17.4%)', 'Engineering (with 13.5%)', 'Environmental Science (with 12.7%)', and 'Social Science (with 11.2%)' are respectively the subject areas that have borrowed the concept of sustainable business models and they all together have published more than half (i.e. 54.8%) of the documents. While during the last three years, from 2016, the focus of these order of subject areas have changed and 'Environmental Science (with 18.6%)', 'Business, Management, and

Accounting (with 16.4%)', 'Social Science (with 14.4%)', and 'Engineering (with 12.3%)' are subject areas that have respectively published the most documents related to sustainable business models which represents a considerable shift of literature of sustainable business models to environmental science and a moderate shift to social sciences.

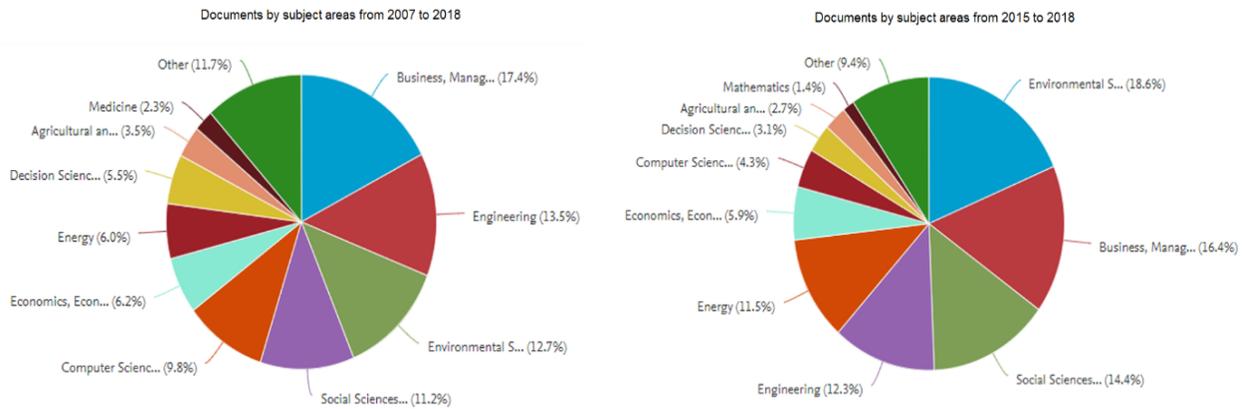

**Figure 4.** Application of sustainable business model in different subject areas.

A precise look at the research on sustainable business models reveals that the research on sustainable business models is more prevalent in the U.S than in other countries. Figure 5 indicates that more than 1250 research related to sustainable business models are carried out in the context of the U.S, from 2007 to 2018. The U.K (with 650 research), China (with 500 research), Germany (with 450 research), and Australia (with 350 research) are respectively countries in which the next highest levels of research is conducted on sustainable business models.

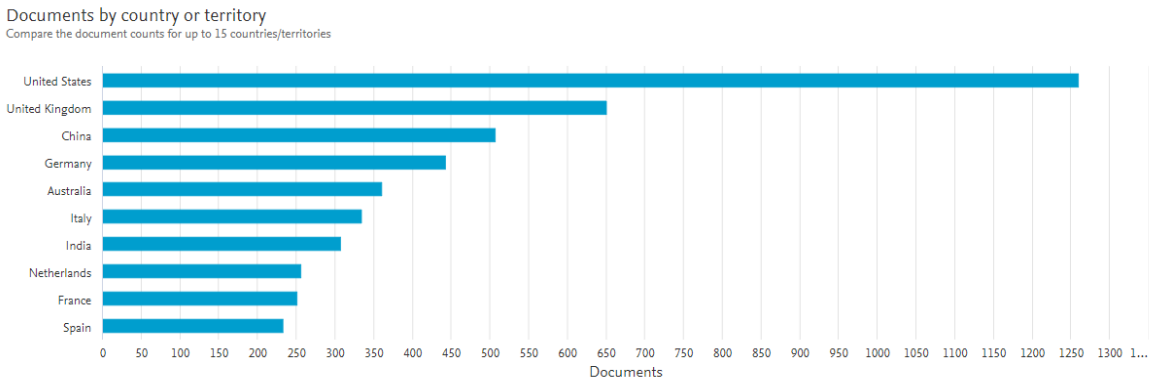

**Figure 5.** Research on sustainable business models in different countries from 2007 to 2018.

Among the documents published in the area of the sustainable business model, 53.8% are original research articles, 29.6% conference papers, 7.9% book chapter, and 3.7% of them are review articles. As figure 6 indicates, the original research article is the most common document published in the area of sustainable business model from 2007 to 2018.

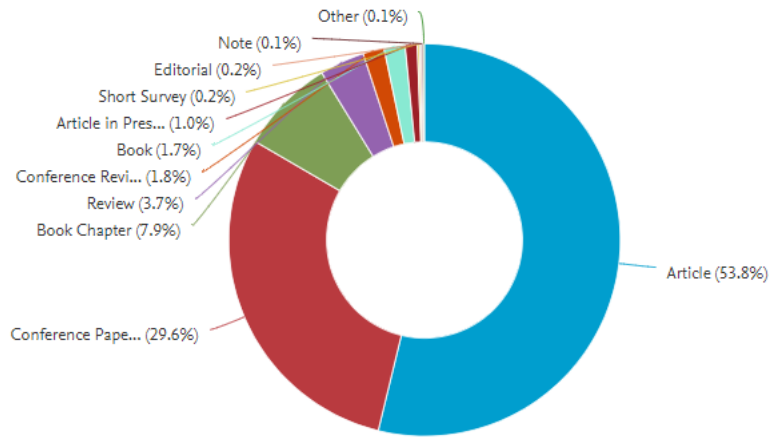

**Figure 6.** Types of documents are published in the area of sustainable business model from 2007 to 2018.

A primary search in the literature of sustainable business models finds that 3688 documents in 27 different subject areas are published. Table 1 constitutes the detail related to these 27 subject areas and the number of articles published in their area utilizing a sustainable business model in their title, abstract or keywords.

**Table 1.** Application of sustainable business models in different subject areas, based on the primary search, from 2016 to 2018.

| Subject area | No of Documents |
|---|---|
| Environmental Science | 687 |
| Business, Management, and Accounting | 603 |
| Social Sciences | 531 |
| Engineering | 454 |
| Energy | 425 |
| Economics, Econometrics and Finance | 216 |
| Computer Science | 158 |
| Decision Sciences | 115 |
| Agricultural and Biological Sciences | 100 |
| Mathematics | 51 |
| Medicine | 50 |
| Arts and Humanities | 49 |
| Earth and Planetary Sciences | 47 |
| Materials Science | 38 |
| Chemical Engineering | 30 |
| Chemistry | 29 |
| Psychology | 25 |
| Physics and Astronomy | 18 |
| Biochemistry, Genetics and Molecular Biology | 17 |
| Multidisciplinary | 10 |
| Health Professions | 8 |
| Pharmacology, Toxicology, and Pharmaceutics | 8 |
| Neuroscience | 6 |
| Nursing | 6 |
| Immunology and Microbiology | 3 |
| Veterinary | 3 |
| Dentistry | 1 |

In the section of materials and methods, the data collection and reviewing process has been explained in detail. In the review section, a picture of the current research on sustainable business models, firstly, is provided and then the application of sustainable business models in different areas are discussed. In the discussion and conclusion section, the findings are articulated and detailed application of the models and recommendations for future research are presented.

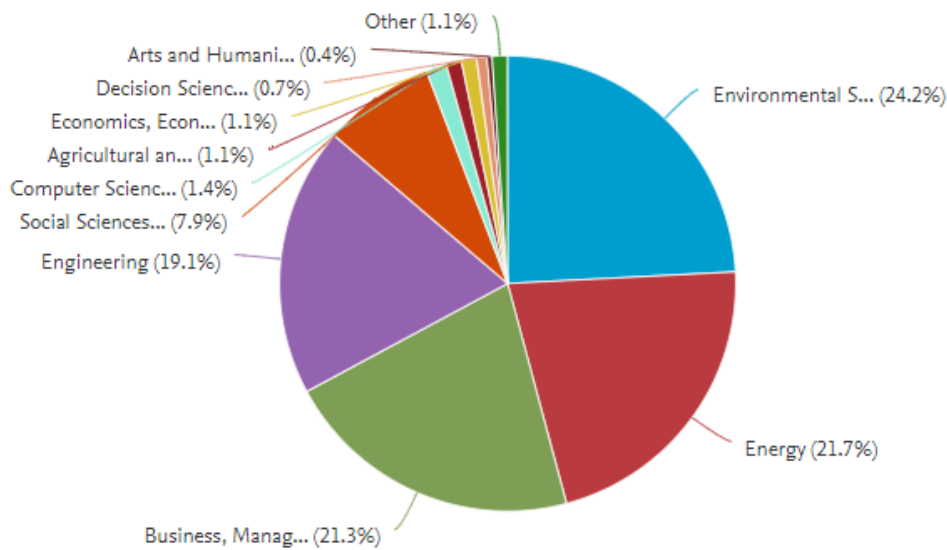

**Figure 7.** The subject areas of the articles are considered for future analysis in this study.

As the above figures indicate Environmental Science, Energy, Business, Management and Accounting, Engineering, and Social Sciences contribute more to the number of documents. Nevertheless, the research methodology classifies the literature into fourteen categories i.e. innovation, management and marketing, entrepreneurship, energy, fashion, healthcare, agri-food, supply chain management, circular economy, developing countries, engineering, construction and real estate, mobility and transportation, and hospitality. Figure 8 illustrates this classification.

Sustainable business models leverage the firms to integrate their economic objectives with their sustainability ambitions in such a way that the benefits of all the stakeholders are achieved simultaneously [22]. Porter and Kramer [23] argue that sustainable business models are sources of competitive advantage in which incorporating sustainable value proposition, value creation and value capturing mechanisms bear economic benefits to the companies. Boons and Lüdeke-Freund [11] count four main characteristics of a sustainable business model distinguishing it from a conventional business model. They believe that the value proposition of sustainable business models is an ecological or social value in accordance with economic value. In the supply chain of sustainable business models' suppliers feel a responsibility towards the focal company's stakeholders as well. Sustainable business models encourage sustainable consumption. Ultimately, Boons and Lüdeke-Freund [11] express that in the design of the financial model of the sustainable business models, in addition to the economic benefits, the company's ecological and social impacts are also considered. Abdelkafi & Täuscher [24] define sustainable business models as tools incorporating sustainability in the firms' value proposition and value creation logic. Per se, sustainable business models not only provide value to their customer but also to the natural environment and society. Geissdoerfer, Bocken, and Hultink [25] consider sustainable business models as a set of the elements in which the interrelation between these elements and their interactions with the stakeholders create, deliver, capture, and exchange sustainable value for its multi-stakeholders.

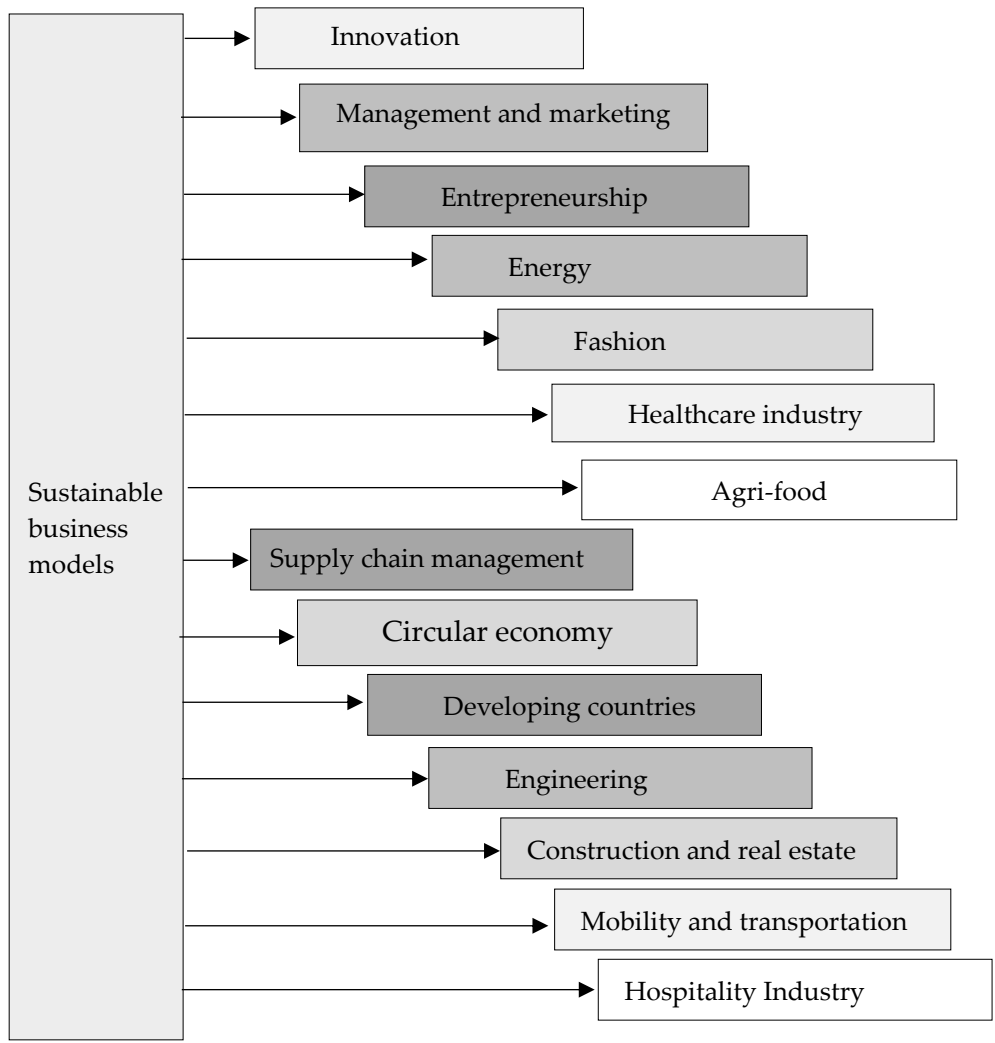

**Figure 8.** Taxonomy of application categories of sustainable business models

Businesses with different characteristics from different industries are aided by sustainable business models to achieve their sustainability ambitions. Besides, many researchers have incorporated this concept with other concepts to provide possible solutions for businesses for sustainable development. Further, this section articulates in detail how sustainable business models are applied in the specific categories of innovation, management, and marketing, entrepreneurship, energy, fashion industry, healthcare industry, agri-food, supply chain management, circular economy, developing countries, construction and engineering, hospitality industry.

*3.1. Innovation*

Much research has been conducted on the common fields of innovation and which mainly strive to propose models, frameworks, or guidelines to elaborate how to innovate a sustainable business model or how to shift a traditional business model to a sustainable business model. Evans *et al.* [15] provide five paradigms for such transformation toward a sustainable business model.

Evans *et al.* [15] articulate that the first step to designing a sustainable business model is to design sustainable value that incorporates economic, social and environmental benefits conceptualized as value forms. According to Evans *et al.* [15], the second step to design a sustainable business model is to create a system of sustainable value flows among multiple stakeholders, including the natural environment and society as primary stakeholders. Generating a value network with a new purpose, design, and governance is the third step toward a sustainable business model. The fourth step to have a sustainable business model is to consider systemically the stakeholder interests and responsibilities for mutual value creation. Finally, internalizing externalities through Product Service System also enables innovation towards sustainable business models.

Geissdoerfer, Bocken and Hultink [26], inspired by design thinking, developed the concept of 'Value Ideation' comprising value ideation, value opportunity selection, and value proposition prototyping. Based on the first principle of Evans *et al.* [15] model, the approach of Geissdoerfer *et al.* [26] to design a sustainable business model is to design a sustainable value proposition in which additional forms of value are created by identifying formerly underserved stakeholders (including society and environment) in the value proposition.

Likewise, "Value Triangle" is a new framework to design that is proposed by Biloslavo, Bagnoli, and Edgar [27]. The VT is a tool that allows a firm to capture economic value from a circular value system in which the value is co-created and co-delivered through the collaboration of the firm with its stakeholders. In other words, the value generated in value triangle is able to meet the benefits of customers (customer value), partners and suppliers (partner value), social actors including environment and future generation (i.e. public value), and the firm itself (captured value).

Oskam, Bossink, and de Man [28] propose the concept 'value shaping' for sustainability-oriented innovations that are able to clarify all the types of financial, social and environmental value that a business creates by interacting with the different networks. They outline that depending on the place of the business model in the life cycle curve, different networks assist the business to design the value. Exploring value refers to the value the firm explores through the existing network and the social network of entrepreneurs. Developing value refers to the value the firms are shaping through engagement of the potential customers. Reframing value refers to the stage in which the feedback from the real customers is utilized to refine the delivered value. Finally, Oskam *et al.* [28] argue that redirecting value refers to shifting from the current value to other value/values due to change in the mindset of the firms; or redirecting value from the direct customers towards the end clients of the products.

Joyce and Paquin [29] provide a novel approach to design a sustainable business model. They propose a Triple Layered Business Model Canvas to meet the economic, social, and environmental benefits in which these three layers respectively explain how the value creation and delivering process satisfy the benefits of business, society, and the environment.

Roman, Liu, and Nyberg [30] propose a three-step approach to design a sustainable business model for progressing toward open access databases in which research data created from universities are accessible to industry for facilitating the open innovation process. Their model comprises three stages of identifying the possible opportunities, recognizing the barriers, and finally designing the model.

**Table 2.** Application of Sustainable business models in the innovation.

| Author/s | Year | Contribution | Methodology | Data Source |
| --- | --- | --- | --- | --- |
| Evans et al [15] | 2017 | Framework | Qualitative | Literature synthesis |
| Geissdoerfer, Bocken, and Hultink [26] | 2016 | Framework | Qualitative | Literature synthesis, expert interviews, and multiple workshops |
| Biloslavo, Bagnoli, and Edgar [27] | 2018 | Framework | Qualitative | Systematic literature review, case study |
| Oskam, Bossink, and de Man [28] | 2018 | Framework | Qualitative | Case study, interview, secondary data |
| Joyce and Paquin [29] | 2016 | Framework | Qualitative | Literature synthesis, secondary data |
| Roman, Liu, and Nyberg [30] | 2016 | Framework | Qualitative | Case study, interview |

By taking a close look at the mentioned research in table 2, it is illustrated that all the authors, utilizing a qualitative research methodology, have tried to provide conceptual frameworks for designing a sustainable business model. Interestingly enough, all the authors have seen the solution in the nature of the "value" the businesses are offering to their users and concepts of value ideation, triangle, and value shaping are the consequence of such perspective. Although they have tried to their theoretical frameworks applying case studies, it sounds necessary for the future research that both the researchers and practitioners provide more empirical evidence to prove the proposed models. On the other hand, despite the fact that offering a value proposition which is able to meet sustainability goals is a logical approach, other innovative approaches imposing the businesses to reconsider the value creation, value delivering and even value capturing processes to meet the sustainability requirements can provide valuable solutions as well – which is missing in the literature.

*3.2. Management and Marketing*

Business models have been considered as tools to implement business strategies. Therefore, it makes sense that the goals of the business models should be aligned with the organizations' goals. Research has proved that designing a business model which can provide sustainability to the society, environment and the business itself requires a prerequisite: providing sustainable values to the society and the environment should be considered in the purpose of the organization.

Stubbs [31] by studying the characteristics the sustainable business models of B Corporations illustrates that social and environmental concerns are embedded in the mission and purpose of B Corporations and the main goal of such corporations is to create positive societal impacts for their stakeholders. He realized that such thinking affected the value propositions, value creation and value delivering of B corporations whereby they align their profit and societal impact. It is worth mentioning that businesses can be certified, by B Lab, a nonprofit organization with offices in all the continents, as B Corps if they have had the highest performance in social and environmental standards, public transparency, and legal accountability.

Morioka, Bolis, Evans, and Carvalho [32] conducted multiple case studies in eleven organizations from diverse sectors, situated in Brazil and in the United Kingdom. They realized that to integrate sustainability into sustainable business model's value creation and delivery system, the organizations should firstly, make a connection between business purpose and employees' values and beliefs, then they should be pro-active and clear engagement in solving sustainability problems.

In addition to aligning the goal of business model with the business itself, the role of decision makers in implementing a business model should not be neglected. Kurucz, Colbert, Lüdeke-Freund, Upward, and Willard [33] explain how relational leadership advances the design and assessment of sustainable business models. According to Kurucz *et al.* [33], relational leadership processes that support strongly sustainable organization management help organizations to address effectively the existing constraints and also to avoid contributing to the tightening of future limits of the biosphere. They articulate that by engaging relational leadership in strongly sustainable business model canvas (SSBMC) and the future-fit business benchmark (f2b2) organizations can define and strive for their sustainability goals. Upward and jones (2016) argue that the strongly sustainable business model canvas demonstrates relational leadership characteristics that support business modeling toward strategic sustainability. Additionally, Kurucz *et al.* [33] explain that the future-fit business benchmark (f2b2) provides a 'fourth benchmark' which defines the ultimate goal of zero negative impact on the socio-ecological system

A business model elucidates how a business makes money through value proposition, value creation and value delivering. The core concept in the business model is "value". The value that the customer is ready to pay for it. Most of the marketing activities are dedicated to diagnosing the customers' needs for providing such value for them. The next stream of research on business model sustainability, in the literature, is to engage the final users in the value proposition process. Engaging the end users in the process of designing value is one of the approaches facilitating the businesses to consider customers' benefits and to design a sustainable business model.

By studying firms that provide energy efficiency products and services Tolkamp, Huijben, Mourik, Verbong, and Bouwknegt [34] found that utilizing a user-centered approach to design a sustainable business model is a key to success of these firms. He realized that firms engage the customers in designing their business model in the form of a four-stage loop including the design of involvement, facilitation of involvement, extraction of lessons learned and finally business model adaptation. In other words, Tolkamp *et al.* [34] claim that identifying and incorporating the customer needs into the firm's value proposition is of utmost importance in designing an effective and sustainable business model.

Baldassarre, Calabretta, Bocken, and Jaskiewicz [35], aided by the principle of user-driven innovation, provide a practical framework to design a sustainable business model through designing a sustainable value proposition. User-driven innovation present solutions to meeting the benefits of society and the business, at the same time through an iterative process in which potential customers are engaged in the design of value proposition.

De Bernardi and Tirabeni [36] perceive that designing a sustainable business model involves designing a community-centered sustainable value proposition. By combining principles from both sustainable business model innovation and user-driven anti-consumption and well-being habits they intended to design a sustainable business model that enhance sustainable and anti-consumption behaviors. They studied the Italian Food Assembly, which is a successful example in the Alternative Food Network (AFN). De Bernardi and Tirabeni [36] found that two main factors have caused Italian Food Assembly industry to implement a sustainable business model: 1) there is a strong knowledge sharing of sustainable consumption behavior among the members and 2) there is an effective distribution of best practices among them also.

**Table 3.** Application of Sustainable business models in management and marketing.

| Author/s | Year | Contribution | Methodology | Data Source |
|---|---|---|---|---|
| Stubbs [31] | 2017 | Design & Process | Qualitative | Interview |
| Morioka et al [32] | 2018 | Framework | Qualitative | Case Study |
| Kurucz et al. [33] | 2017 | Conceptual model | Qualitative | Literature synthesis |

| Tolkamp et al. [34] | 2018 | Design & Process | Qualitative | Interview |
| Baldassarre et al. [35] | 2017 | Framework | Qualitative | Literature synthesis, expert interviews, and multiple workshops |
| De Bernardi and Tirabeni [36] | 2018 | Design & Process | Qualitative | Case Study, depth interviews, participant observation, focus groups, and document analysis. |

*3.3. Entrepreneurship*

Davies and Chambers [37] argue that sustainable entrepreneurs encounter hybrid tensions when they focus on creating economic value whilst increasing social or environmental value. They argue that conflicts among different value capturing processes lead to business instability, and a business model innovation is a solution to eradicate the conflict. Gasbarro, Rizzi, and Frey [38] provide empirical insights on how sustainable entrepreneurs cope with regulative, normative and cultural-cognitive issues to increase institutions' legitimacy by developing a sustainable business model. They articulate that institutional entrepreneurs (SIEs) design innovative business models by engaging the final customers and strategic partnerships in developing innovative value propositions process to, firstly, increase the benefit of innovative sustainable business models, secondly, to imitate the possible conflicts, and ultimately to change industry norms and social beliefs and cultural-cognitive barriers in a value proposition to increase legitimacy within the normative and cultural-cognitive institutions.

Khalid, Hassam, and Ahmad [39] consider the entrepreneurial action theory as an alternative to entrepreneurship theory since it has an important role in the sustainable business innovation model. Significant knowledge derived from entrepreneurial action provides a better understanding of how to develop and establish sustainability-innovation ventures. Whilst, Neumeyer, and Santos [40] reveal that although the networks of sustainable entrepreneurial ventures are more densely connected in comparison to conventional entrepreneurs, sustainable entrepreneurs are underrepresented, in the Southeast United States. de Lange [41] also provides empirical evidence illustrating that the investors are reluctant to invest in sustainable start-ups, particularly those are environmentally sustainable. On the other hand, de Lange [41] also illustrate that the investors are attracted to invest in the start-ups in the sustainable national context.

**Table 4.** Application of Sustainable business models in the entrepreneurship.

| Author/s | Year | Contribution | Methodology | Data Source |
| --- | --- | --- | --- | --- |
| Davies and Chambers [37] | 2018 | Theoretical and empirical evidence | Quantitative | Multiple case study, interview |
| Gasbarro et al. [38] | 2018 | Empirical evidence | Qualitative | Interviews and archive data |
| Khalid et al. [39] | 2016 | Framework | Qualitative | Literature synthesis |
| Neumeyer and Santos [40] | 2018 | Empirical evidence | Quantitative | Literature synthesis, interview, secondary data |
| de Lange [41] | 2017 | Empirical evidence | Quantitative | Secondary data |

*3.4. Energy*

One of the objectives of a sustainable business model is to eliminate (or at least minimize) the harmful effect of the businesses on the environment. Many approaches are provided in the literature for the businesses to reach this sustainable goal. Management of the resource and energy is of utmost importance in meeting the sustainability goals. Moschetti, Brattebø, Skeie, and Lien [42] propose an analytic process based on the execution of quantitative sustainability analyses, to transition from a traditional focus of business models on economic value and customers toward proposing, creating, and capturing sustainable values for the environment and the society. Sousa-Zomer and Cauchick Miguel [43] investigate how such a sustainable business model can support technological innovations such as decentralized approaches for water quality and quantity improvements in urban areas. Their research revealed that having a sustainable business model through close integration with customers improve consumers' acceptance, risk perception, and confidence in decentralized approaches. Tah and Abanda [44] illustrate that the internet of things has presented many opportunities to reduce the consumption of energy and carbon emissions by introducing concepts such as intelligent buildings and smart cities.

Zhang, Guo, Gu, and Gu [45] propose a framework which assists the decision makers to develop sustainable business models for high energy-consuming equipment (HECE). Aided by Product-Service System (PSS), they suggest a decision-making support tool for developing PSS of HECE. In their opinion, in a sustainable business model, the benefits of all the stakeholders are considered. The illustrated that extra economic benefits impose more burdens and higher risk to the environment. And developing PSS is not always sustainable as in gas supply service, for instance, it would lead to extra economic and environmental burdens due to frequent transportation.

Rossignoli and Lionzo [46] provides empirical evidence of new forms of interdependencies arising within partnership networks that drive businesses in the energy sector to have a sustainable business model. He believes that a network induces its contributors to expand their definition of value and requires them to create value for both companies and society as the main objective of their business model. According to Rossignoli and Lionzo [46], the new links among participants of a network create new approaches for capturing value and assist them to solve the concerns related to resource dependency, which is achieving sustainability goals.

Nichifor [47] conducted research to compare the current sustainable business models of the current firms in the wind and solar energy sector in Romania. She found that Both sectors have encountered egregious changes in last two years due to the changes in supporting schemes by the European Union and the government that affect renewable energy markets. In addition, Nichifer [47] found the investors in the wind energy sector found a pessimistic outlook of future investments due to legal instability that made them reduce the wind projects.

**Table 5.** Application of Sustainable business models in the energy section.

| Author/s | Year | Contribution | Methodology | Data Source |
| --- | --- | --- | --- | --- |
| Moschetti et al. [42] | 2018 | Model | Quantitative | Literature synthesis and case study |
| Sousa-Zomer and Cauchick Miguel [43] | 2018 | Design & process | Qualitative | Case study |
| Tah and Abanda [44] | 2017 | Empirical evidence | Qualitative | Literature synthesis and case study |
| Zhang et al. [45] | 2018 | Framework | Qualitative | Literature synthesis and case study |
| Rossignoli and Lionzo [46] | 2018 | Empirical evidence | Qualitative | Case study, interview, questionnaire |
| Nichifer [47] | 2015 | Empirical evidence | Qualitative | Case study, interview, questionnaire |

*3.5. Fashion*

Pal and Gander [48] also believe that the traditional business models in the fashion industry produce highly negative outcomes for the environment through high water usage, chemical pollutions, and incineration or landfill of large amounts of unsold stock. Ciasullo, Cardinali, and Cosimato [49] claim that the fashion industry is unsustainable as active companies in this industry imposed many economic, social and environmental burdens. Therefore, researchers have tried to provide tools and approaches to design a sustainable business model which is able to cope with the social and environmental issues in the fashion industry. Kozlowski, Searcy, and Bardecki [50], for instance, develop a new design tool, called the reDesign canvas, to assist sustainable designers in the fashion industry. They propose a business model canvas with 12 building blocks ensuring the entrepreneurs build a sustainable fashion brand. Hirscher, Niinimäki, and Joyner Armstrong [51] aided by social manufacturing theory strived to design a more sustainable innovative value in design and manufacturing of fashion. They use do-it-yourself (DIY), do-it-together [52] design strategies which users turned into the value creators to develop a sustainable business model. The DIY strategy allows consumers to be both the designer and the producer of their own garment. The producer provides them so-called DIY kits that contain materials and instructions. Whilst, Hirscher, Niinimäki, and Joyner Armstrong [51] argue that DIT is workshops that enable the users to design and build the garments together while they use one another skills and knowledge.

Slow fashion is an approach aimed at intensifying sustainability in the fashion industry. Jung and Jin [53] conducted research to investigate the profitability of this approach in the fashion industry. Customer value creation framework, which is one of the slow fashion solutions, refers to creating perceived customer value. They provide empirical evidence that involving the customers in the value creation process increases their intention to pay a price premium for slow fashion products. Jung and Jin [53] found that creating customer value for slow fashion positively affects consumers' purchase intentions which can secure an economically sustainable business model, while continuously ameliorating environmental and social sustainability.

Pal and Gander [48] argue that incongruence of fashion customers' values with the value propositions and the barriers in the transition of supply chain toward a slowing and a closing loop of resources is detrimental to developing a sustainable business model in the fashion industry. [48] They believe that development of a business model should be considered as a system for creating value for the customer and environment and also capturing value for the firm so that the firms can replace the dominant, unsustainable models with sustainable business models in the fashion industry.

Table 6. Application of Sustainable business models in the fashion industry.

| Author/s | Year | Contribution | Methodology | Data Source |
|---|---|---|---|---|
| Pal and Gander [48] | 2018 | Theoretical Evidence | Qualitative | Literature synthesis |
| Kozlowski *et al.* [50] | 2018 | Framework | Qualitative | Literature synthesis, participatory action research (PAR), and interviews |
| Hirscher *et al.* [51] | 2018 | Framework | Qualitative | Literature synthesis, workshop, and interview |
| Jung and Jin [53] | 2016 | Empirical evidence | Quantitative | Questionnaire |

*3.6. Healthcare Industry*

Nikou and Bouwman [[54] conduct a systematic literature review based on a business model ontology to find the applications of mobile technology and devices in the healthcare industry. Their findings illustrate that in order for Mobile Technology to contribute to the design of sustainable business models in the healthcare industry, non-technological business model components such as value proposition, organizing, and revenue models should be considered rather than focusing on the service platforms. In other words, to design a sustainable business model in the healthcare industry by utilizing Mobile Technology, value propositions should be designed based on the customer's values to provide social benefits and the value capture processes should be designed to provide economic benefits. Merchant, Ward, and Mueller [55] claim that utilizing Telemedicine (also known as telehealth) is a tool that provides sustainability to hospitals. According to Merchant, Ward, and Mueller [55], Telemedicine provides solutions to design value propositions to develop a sustainable business model. Their results disclose that, although, hub hospitals are more responsible for the design of sustainable business models in comparison to the spoke hospitals in the U.S., both hub and bespoke hospitals pointed out that telemedicine helps them to meet their mission, facilitates access, keeps lower-acuity patients closer to home, and helps head off competition. However, Anwar and Prasad [56] argue that although telemedicine has presented many solutions for developing sustainable business models in the healthcare industry, the adoption of such technology has turned into the utmost importance. Because evolution and sometimes revolution in this technology has made it hard for the users to get used to it. Anwar and Prasad [56] recommend a continuous eHealth literacy for, firstly, facilitating the transition era and secondly, the development of new business models in which the users' involvement and motivation and also revenue generation have been considered. They express that telemedicine services should be user- friendly and sustainable which are able to integrate all stakeholders' benefits in one system.

Table 7. Application of Sustainable business models in the healthcare industry.

| Author/s | Year | Contribution | Methodology | Data Source |
| --- | --- | --- | --- | --- |
| Nikou and Bouwman [54] | 2017 | Theoretical evidence | Qualitative | Literature synthesis |
| Merchant, Ward, and Mueller [55] | 2015 | Theoretical & empirical evidence | Qualitative | Literature synthesis, secondary data, and interviews |
| Anwar and Prasad [56] | 2018 | Framework | Quantitative | Literature synthesis |

*3.7. Agri-food*

Research interest in providing sustainable solutions for developing business models in the agri-food sector has increased in these years [57]. Franceschelli, Santoro, and Candelo [58] argue that development of sustainable business model innovation within the food industry, especially for start-ups, is of utmost importance because the industry is itself linked with nature and human respect. Franceschelli *et al.* [58], utilize a business model canvas to design an innovative sustainable business model for food start-ups. Barth, Ulvenblad, and Ulvenblad [57] by conducting a systematic literature review propose a conceptual framework for sustainable business model innovation in the agri-food sector which can meet the challenges encountered in taking a sustainability perspective.

Lee and Slocum [59] study the event organizers who plan the events for food and beverage providers. They provide empirical evidence of event organizers having a sustainable business model and organize the events for the local foods. Although they have contractual flexibility to select foods, there is a willingness to pay a price-premium for local products. Lee and Slocum [59] also show that the meeting/event attendees have not considered themselves sustainable yet and there is a need to increase the knowledge of and the benefit of local foods (which are organic and harmless for the environment) to enhance the attendees' knowledge about sustainability.

Robinson, Cloutier, and Eakin [60] prove that the landscaping enterprises have a sustainable business model, thereby provide multifunctional edible landscapes in the cities, have a greater range of value propositions and revenue streams resulting in increasing their competitive advantage. They express that these enterprises can have the potential value creation of edible landscaping ranged between $3.9 and $66 billion and that positive return on investment (ROI) could be achieved within one to five years.

Table 8. Application of Sustainable business models in the agri-food section.

| Author/s | Year | Contribution | Methodology | Data Source |
| --- | --- | --- | --- | --- |

| Barth et al. [57] | 2017 | Framework | Qualitative | Systematic literature review |
| Franceschelli et al. [58] | 2018 | Theoretical Evidence | Qualitative | Case study, secondary data, and interviews |
| Lee and Slocum [59] | 2015 | Empirical Evidence | Quantitative | Questionnaire |
| Robinson et al. [60] | 2017 | Empirical Evidence | Qualitative | Interview, and GIS landscape analysis |

*3.8. Supply Chain Management*

Supply chain management is another sector that has borrowed the concept of a sustainable business model as a possible solution to meet sustainable development. The objective of sustainability is to address environmental and socio-economic issues in the long term [61]. Ray and Mondal [62] provide evidence to illustrate that collaboration is better than the competition to keep staying in the market. They argue that collaboration among firms within a closed-loop supply chain (CLSC), which its aim is to minimize the waste by inputting the returned used products or parts of the products into another manufacturing process. Therefore, Ray and Mondal [62] propose a collaborative business model and mechanism for collaborative business strategies in a CLSC. Witjes and Lozano [61] also provide evidence that collaboration is crucial to develop sustainable business models for supply chain management. They believe that collaboration between procurers and suppliers in the procurement process mitigates the use of raw material and waste generation through the development of sustainable business models. Witjes and Lozano [61] declare that in a collaboration business model, suppliers and procurers gain experience in the collaboration process to improve circular economy objectives and to secure economic benefits for both parties, by developing sustainable business models that lead to reductions in raw material utilization and waste generation.

Geissdoerfer, Morioka, de Carvalho, and Evans [63] inspired by the circular business model concept and circular supply chain management concepts strive to design a sustainable framework to provide solutions for sustainable supply chain management. They disclose that the circular business model provides a different solution for different loops: closing loops, slowing loops, intensifying loops, narrowing loops, and dematerializing loops.

Brennan and Tennant [64] conducted a comparative case study to find out how to resolve trade-offs in sustainable supply chain management. They realized that for the transition from a traditional supply network toward sustainable supply network, business model innovation requires the creation of sustainable values and resolving trade-offs. They illustrate that network-centric business model innovation provides sustainable solutions for the trade-off between economic and environmental benefit through the prioritization of sustainability-related 'cultural' resources.

**Table 9.** Application of Sustainable business models in supply chain management.

| Author/s | Year | Contribution | Methodology | Data Source |
|---|---|---|---|---|
| Ray and Mondal [62] | 2017 | Framework | Qualitative | Systematic literature review |
| Witjes and Lozano [61] | 2016 | Theoretical Evidence | Qualitative | Literature synthesis |
| Geissdoerfer *et al.* [63] | 2018 | Framework | Qualitative | Literature synthesis, case study, interviews |
| Brennan and Tennant [64] | 2018 | Empirical Evidence | Qualitative | Case study and Interview |

3.9. Circular Economy

The circular economy in the literature is widely considered as a tool to implement and design a sustainable business model in the different sectors in response to currently unsustainable trajectories. As it is shown in table 9, Witjes and Lozano [61] and Geissdoerfer *et al.* [63] utilized this concept to design a sustainable business model for the area of supply chain management. In this section, the other articles that have benefited from the Circular Economy for designing a sustainable business model are compiled and discussed.

Heyes, Sharmina, Mendoza, Gallego-Schmid, and Azapagic [65] applied Backcasting and Eco-design for the Circular Economy (BECE) framework to identify how ICT firms diagnose circular business model innovations. Since BECE is designed for the product-oriented firms Heyes *et al.* [65], by shifting the focus to a user-centered eco-design, design circular economy models that put the customer satisfaction in priority.

Todeschini, Cortimiglia, Callegaro-de-Menezes, and Ghezzi [66] by synthesizing the current literature have developed an innovative circular business model in which the value propositions are sustainable and reduce environmental impacts. By conducting eight case studies on innovative fashion startups, they identify the concept of 'born sustainable' which assist the entrepreneurs to design sustainable value propositions to accomplish the circular economy objectives.

However, Stål and Corvellec [67] provide empirical evidence, based on seven case studies in Sweden, that the businesses are pro-actively looking for solutions to increase institutional demands for circularity to meet their own economic interests (rather social and environmental benefits). Their findings show that the businesses buffer their business model and their value proposition from emerging demands (demands for sustainability) by outsourcing and internal separation.

**Table 10.** Application of Sustainable business models in the circular economy.

| Author/s | Year | Contribution | Methodology | Data Source |
|---|---|---|---|---|
| Heyes *et al.* [65] | 2018 | Framework | Qualitative | Literature synthesis, case study, workshop |
| Todeschini *et al.* [66] | 2017 | Framework | Qualitative | Literature synthesis, case study, interviews |
| Stål and Corvellec [67] | 2018 | Empirical Evidence | Qualitative | Literature synthesis, case study, interview |

3.10. Developing Countries

Research conducted on sustainable business models in developing countries mainly addresses the bottom of the pyramid (BOP) context, where there is a paucity of resources and population suffer from poverty. Bottom of the pyramid refers to the global poor who are in extreme poverty and are unable to meet basic needs [68], most of whom live in developing countries. According to the World Bank reports, 2.7 billion people, which are around half the global population, have less than $2 a day income [68].

Bittencourt Marconatto, Barin-Cruz, Pozzebon, and Poitras [69] provide evidence illustrating that the Brazilian government facilitates the transition toward a sustainable business model by providing strategic and shared value opportunities. By studying the Ecoelce project, they articulate how to design a sustainable business model in the BOP context of Northeastern Brazil. In the Ecoelce project, Bittencourt Marconatto et al. [69] they intended to stimulate low-income customers to exchange recyclable waste for discounts on electrical bills. To describe a sustainable business model, Bittencourt Marconatto et al. [69] have considered value proposition, supply chain, and value capture as the main components of a sustainable business model. Where the value proposition refers to the value that the project provides to their community and the supply chain points out the necessary actions for creating and delivering value to the final users and finally the value capture explains that how the project can make money through these value creation and value delivering process [69]. Bittencourt Marconatto et al. [69] consider discounts in the electricity bills so as to encourage to exchange the wastes. Besides, they have tried to minimize the solid wasted improperly disposed of in low-income communities. Reduce energy theft, level of client's insolvency and illegal connection are of the value propositions they have considered for the low-income communities in their project. Bittencourt Marconatto et al. [69] utilized the closed loop supply chain principles to make longer the life cycle of the wastes and they sent the waste to a third party rather than sending them directly to the recycling companies. They changed the value capture model in a way that the price per material was adjusted by back-office software to avoid a potential loss for both parties (the third-party company and the recycle company) while the recycle company can be ensured about longer incomes. Dembek, York, and Singh [70] provide nine individual business models addressing poverty through studying 55 organizations in Indonesia and the Philippines. They realized that the businesses in these two regions have three different types of business models: 1) delivery models that provide products and services to the BoP communities, 2) sourcing models that create products and services by the members of BoP communities and deliver them to the non-BoP communities and the international markets,and 3) reorganizing models that modify existing systems and ways of life to benefit BoP communities. Goyal, Sergi, and Kapoor [71] provide strategic solutions for social enterprises to develop a sustainable business model which can meet the underserved needs of the BoP segment in India. They propose a practical framework for creating a sustainable, scalable and socially relevant ecosystem. Their framework constitutes a proposition which claims: 1) demographic variables of a BoP community affects the social enterprises performance, 2) field-based experimentation, innovation and prototyping generate customized values for the BoP communities, 3) engaging the local people (the users) in the value creation process to positively affect the social enterprise's performance, 4) hybrid organizational setup can align the social benefits with organizational economic benefits, 5) social marketing, the product quality and service support can get the trust and acceptance of a BoP community, 6) providing need-based customized end to end solutions leads to trust and acceptance also, 7) the brick and mortar delivery channels and local engagement-based hub and spokesmodels positively affects the social enterprise's accessibility and availability, 8) collaboration with stakeholders positively affects market reach and acceptance of the social enterprise, and a fine-tuning the business model positively affects the socio-economic impact and the social enterprise's performance. Palomares-Aguirre, Barnett, Layrisse, and Husted [72] study business models of three firms that provide affordable housing for very poor people in Mexico. Their finding reveals that community engagement and government collaboration are very important in creating and delivering a sustainable value so as to better serve the BoP.

**Table 11.** Application of Sustainable business models in developing countries.

| Author/s | Year | Contribution | Methodology | Data Source |
|---|---|---|---|---|
| Bittencourt Marconatto *et al.* [69] | 2016 | Empirical Evidence | Qualitative | Case stud, observations, interviews and secondary data |
| Dembek *et al.* [70] | 2018 | Framework | Qualitative | Primary and secondary data |
| Goyal *et al.* [71] | 2017 | Framework | Qualitative | Interviews and secondary data |
| Palomares-Aguirre *et al.* [72] | 2018 | Empirical Evidence | Qualitative | Literature synthesis, case study, interview |

3.11. Engineering

Construction, the biggest industry in the developed world, has the greatest environmental impact [73] as well as economic and social consequences [42]. However, Selberherr [74] claims that sustainable buildings bear many potential benefits for service providers and society. Selberherr [74] proposes strategies for the players in the construction sector to proactively contribute to the sustainable development of society. She recommends designing a sustainable business model which is aimed at cooperatively optimizing buildings and infrastructures and taking the responsibility for the operating phase via guarantees.

Wasiluk [75], based on the finding resulted from a case study of the Australian property and construction sector, proposes that businesses in lieu of justifying the business case for sustainability they should concentrate on understanding how to mobilize their intellectual capital to enhance an ecologically sustainable and socially equitable enterprise. Indeed, she considers the intellectual capital as a mediator sophisticating sustainable value proposition for the Australian property and construction sector.

Boo, Dallamaggiore, Dunphy, and Morrissey [73] argue that there are approximately 190 million buildings in Europe which were built before energy efficiency was a common issue in construction. They consider innovative business models (IBM) as a solution to provide sustainability in the energy efficient building market. Boo, Dallamaggiore, Dunphy, and Morrissey [73] propose sustainable business models ensuring long-lasting change in the energy efficient building market. They believe that the co-evolution of business models with both the wider energy system and the natural environment is crucial for the development of a sustainable business model.

**Table 12.** Application of Sustainable business models in construction and engineering.

| Author/s | Year | Contribution | Methodology | Data Source |
| --- | --- | --- | --- | --- |
| Selberherr [74] | 2015 | Theoretical Evidence | Qualitative | Literature synthesis |
| Wasiluk [75] | 2013 | Empirical Evidence | Qualitative | Case study, interview |
| Boo *et al.* [73] | 2016 | Framework | Qualitative | Literature synthesis |

3.12. Construction and real estate

One of the serious challenges today's cities are confronted is to design and manage a city sustainably [76]. Sustainable urban development is a field has emerged to address such concerns [77]. International organizations, governmental bodies, and academic institutions have focused on different approaches to evaluating urban performance to recognize the problems and design policies and strategies [78]. Although the research on sustainable cities mainly provides solutions to address the environmental issues, there are research has focused on social sustainability paradigms as well [79].

Rajakallio, Ristimäki, Andelin, and Junnila [80] believe that the business model of the firms that are active in the real estate and construction sector are tied with one another as these should be seen as a network creating and delivering value to their client. Besides, they argue that to develop and construct real estate sustainably, the clients play a vital role as the actual users of the buildings are often tenants who appraise the quality of the buildings. They also note that the buildings are traded in the investment markets, where the value is evaluated by the investor. Therefore, they recommend a joint alignment of design themes in which the stakeholders have the ability to maximize their own private benefits. However, this finding is in contrast with the findings of Bos-de Vos, Volker, and Wamelink [81] who realized that engaging the final users in the value creation step and the designing stage reduces the bargaining power of the firms which will finally lead in a reduction in their economic benefits.

Rivière, Verges, Dimou, and Garde [82] investigate how they can design a network business model for Beauséjour sustainable town project which is a project to build a tropical sustainable city in Reunion Island - a small French island located in the Indian Ocean. The main challenge of the tropical sustainable city is to cope with classic urban issues, environmental concerns, and advocacy planning, simultaneously. They were looking for a network business model that can explain how interaction among the developer-contractor, real-estate developers (housing and services sector) and assets and property management enable them to build a sustainable tropical sustainable city. Their finding reveals that in order to design a sustainable business model, mix environmental issues with the urban project objectives. Rivière *et al.* [82] argue that urban objectives should be translated into an environmental and bioclimatic sensitive design. Furthermore, advocacy planning will allow the inhabitants to take care of the plant heritage and contribute to the project.

The sharing economy, which is one of the principles of the circular economy, is a solution to reach sustainability. Coworking spaces also result from the concept of sharing economy. Waters-Lynch and Potts [83] believe that the research on coworking spaces is disclosing differentiated product niches in the urban office rental market. Waters-Lynch and Potts [83] provide a model considering coworking spaces as 'social economy Schelling points' within the evolving landscape of new spaces for urban production. According to this proposed model the coworking spaces entrepreneurially establish focal points for tacit coordination between niche actors who expect to find each other at such locations to cooperate on joint projects.

According to Yan, Wang, Quan, Wu, and Zhao [84] urban sustainable development efficiency (USDE) explains how efficient an urban system is in meeting the human welfare and resources and the environmental input. Yan *et al.* [84] present a framework to evaluate the performance of urban sustainable development in utilizing natural resource limitations and meeting human welfare needs. Their model constitutes 11 specific indicators including water consumption, area of construction land, fossil energy consumption, life expectancy, government spending on education, living area, Engel's coefficient (Engel law refers to the negative relationship between the income and the proportion of income allocated for food), percent of GDP contributed to others, green land area, days of fairly good air quality, sewage discharge.

Song [85] explains all the three main pillars of a sustainable business model (i.e. economic, environmental, and social benefits) for urban sustainable development. Song [85] presents a theoretical background on how to set up an eco-city for urban sustainable development. He argues that resource consumption is a very important element for sustainable development of the eco-cities. Song [85] claims that resource-saving and environment-friendly industries, reduced resource consumption, and reduced unit GDP resource consumption are the main pillars in the ecological city construction. Song [85] also debates that the construction of eco-cities should be socially sustainable in order to reach the sustainable development objectives of urban areas. He believes that the objective of eco-cities is not only to protect the contemporary human rights but also to ensure the development of human rights for the next generations. In the case of eco-cities, economic development includes the development of ecological agriculture, industry, and services. Eco-efficiency and ecological benefits of economic development are very important in the construction of eco-city [85].

As mentioned above, the main objective of sustainable business models is to provide win-win solutions to meet the economic, social and environmental benefits at the same time. The aim of the study of Li et al. [86] was to develop indicators and an assessment method by which to evaluate the status of urban sustainable development. Li et al. [86] also developed a Full Permutation Polygon Synthetic Indicator method for evaluation of the capacity for urban sustainable development at different times for the next two decades. They developed a system of 52 indicators of urban sustainable development. These 52 indicators evaluate four main dimensions of economic growth and efficiency, ecological and infrastructural construction, environmental protection, social and welfare progress, in a higher level of evaluation.

Many approaches have been proposed in the literature for developing sustainable construction so that they are both eco and socio friendly. At the same time determining the economic value of such constructions is of utmost importance. Zavadskas *et al*. [87] developed a neutrosophic Multi-Attribute Market Value Assessment (MAMVA) method to determine the real market value of property incorporating sustainability aspects. The MAMVA by utilizing the multiple criteria analysis evaluates sustainable buildings considering the vagueness aspects of the initial information. They argue that this method can assist property sellers, brokers, buyers, and lenders on regional, national and global levels also.

**Table 13.** Application of Sustainable business models in construction and real estate.

| Source | year | Contribution | Methodology | Data Source |
|---|---|---|---|---|
| Rajakallio *et al.* [80] | 2017 | Empirical Evidence | Qualitative | Literature synthesis, case study, interviews |
| Rivière [82] | 2013 | Empirical Evidences | Qualitative | Case study |
| Waters-Lynch and Potts [83] | 2017 | Empirical and Theoretical Evidence | Qualitative | Primary Ethnography data |
| Yan et al. [84] | 2018 | Framework | Quantitative | A literature synthesis and Secondary data |
| Song [85] | 2011 | Theoretical Evidences | Qualitative | Literature synthesis |
| Li et al. [86] | 2009 | Framework and Empirical Evidence | Qualitative | Literature synthesis, Case study, and secondary data |
| Zavadskas [87] | 2017 | Framework and Empirical Evidence | Quantitative | Literature synthesis, Case study |

3.13. Mobility and transportation

One of the most important challenges toward global sustainable development is an Urban transformation as mobility sector has a great potential to reduce the carbon emission [88]. Recently disruptive business model innovation has emerged such as app-based smart-sharing systems such as car-pooling, expanded electric vehicle use, and bike-sharing. Such sharing mobility business models plus low-carbon transport modes in cities are able to lead urban mobility toward sustainability. Ma, Rong, Mangalagiu Thornton, and Zhu [88] study the relationship between social-ecological innovation in the sharing economy and urban sustainable development. Conducting three business cases in the emerging sharing mobility sector –ride-sharing, EV-sharing, and bike-sharing - in Shanghai, China, illustrate that there is a strong co-evolution mechanism between the transition towards a sustainable city and the business ecosystem innovation towards a greener and smarter transport. Ma *et al.* [88] believe that the disruptive innovation of the sharing economy is the common area linking this interaction.

Mozos-Blanco, Pozo-Menéndez, Arce-Ruiz, and Baucells-Aletà [89] analyze the effectiveness of Sustainable Urban Mobility Plans (SUMPs) in 38 of the Spanish Network of Smart Cities, in 2018. The sharing economy principles, which is one of the approaches of implementation of the circular economy, is the main criteria considered in the SUMPs in the Spanish Network of Smart Cities. Their finding discloses that although most mobility plans tend to improve pedestrian and cycling mobility, there is a need to provide the required software and hardware infrastructures. Car-pooling or car-sharing does not have any remarkable share in transportation in Spain, therefore parking regulation is another criterion that has emerged in SUMPs by restricting the presence of parking areas around office buildings and residential areas. Mozos-Blanco *et al.* [89] argue that the criterion which has been considered in SUMPs to meet the social and environmental benefits objectives, which are part of the principles of implementation of the sustainable business model in urban development, is the reduction of air and noise pollution and establishing urban green spaces.

Lyons [90] provides theoretical evidence about smart urban mobility and believes that the terms of smart and sustainable are strongly tied with each other. Lyons [90] argues that for development of smart cities only technological development is not adequate, but also there is a need for sociotechnical development to reach the smart urban mobility. He also believes that the appreciation of people's lifestyles, constraints, needs, desires and behaviors as well as the practices of businesses are of the main requirements of achieving the smart urban mobility objectives. In addition, Lyons [90] debates that ICT plays a vital role in supporting how society connects and it can determine address effectiveness and attractiveness of mobility for the user.

Nowickaa [91] believes that sustainable mobility integrates the realization of the needs of stakeholders by using remote access to the properties of desired goods and services. From his point of view, utilizing a cloud computing model in mobility promotes sustainable mobility, minimizes the negative impact on the environment, and increases the social and economic benefits. Nowickaa [91] argues that the use of cloud computing models reduces the total cost of provided services for residents; provides agility, flexibility, and elasticity; quick and cost-efficient reaction to less-predictable events and changing stakeholders' requirements; globally accessible services, easy and fast implementation and strong support for sustainable development.

Köse *et al.* [92] believe that sustainable manufacturing provides competitive advantages to the companies and despite the pressure of stakeholders such as customers, investors, competitors, interest groups and local municipalities, companies voluntarily over-comply with social and environmental norms to take advantage of being sustainable. Köse *et al.* [92], by studying the incentives in the urban mobility to apply sustainable approaches, disclose that the common incentives and strategies of overcompliance drives public and private initiatives toward a sharing economy. They realize that companies design their strategies under the effects of overcompliance with social and environmental aspects to improve sustainability. Köse *et al.* [92] suggest that differentiating existing product lines in favor of sustainability (e.g. electric cars, bamboo bicycles) or by introducing new products that can offer even higher sustainability (e.g. the SUW) can be the possible strategies for the manufacturing industry to overcompliance with social and environmental requirements

Zawieska and Pieriegud [93] consider smart cities and sustainable transportation particularly with regard to the reduction of $CO_2$ emissions. They believe that meeting the reduction targets set by the European Union 2011 White Paper on Transport will be very challenging and it is needed for a profound transformation of transport and energy sectors. Zawieska and Pieriegud [93] also believe that smart city solutions can mitigate transport $CO_2$ emissions and meet reduction goals.

**Table 14.** Application of Sustainable business models in Mobility.

| Source | year | Contribution | Methodology | Data Source |
| --- | --- | --- | --- | --- |
| Ma *et al.* [88] | 2018 | Empirical Evidence | Qualitative | Literature synthesis, case study, secondary data, interviews, surveys, stakeholder workshops |
| Mozos-Blanco *et al.* [89] | 2018 | Empirical Evidence | Qualitative | Case study, secondary data |
| Lyons [90] | 2016 | Theoretical Evidence | Qualitative | Literature synthesis |
| Nowickaa [91] | 2016 | Theoretical Evidence | Qualitative | Literature synthesis |
| Köse *et al.* [92] | 2016 | Theoretical Evidence | Qualitative | Literature synthesis and secondary data |
| Zawieska and Pieriegud [93] | 2018 | Empirical Evidence | Qualitative | Literature synthesis, Case study, and the primary date |

3.14. Hospitality Industry

Hotels are counted as one of the most important sectors of the hospitality industry which are affected by the sustainability movement. The research on the common field of the business model sustainability and the hospitality industry is still in the infancy stage. As most of the studies have tried to investigate the sustainability level of the hotels rather than providing solutions for the development of a sustainable business model in this industry.

Buffa, Franch, and Rizio [94] utilizing a quantitative approach provide empirical evidence that medium-sized hotel enterprises (SMHEs) in Trentino, a traditional tourist destination in the Italian Alpine

Arc, apply sustainable business models. They argue that these SMHEs adopted three different sets of environmental management practices (EMPs) to accomplish their sustainability goals of their business models. Utilizing the factor analysis, Buffa, Franch, and Rizio [94] found out the first group of the practices for implementation of a sustainable business model includes variables that determine the firms' strategies related to environmental protection. These variables are waste, green events, green reporting, green marketing, environmental monitoring, environmental objectives. The second group of the variables is about the alternative heating solutions which are biomass and multi-fuel boilers. The third practices they found was about the variables for the management of the structural changes to improve energy efficiency (renewables, insulation).

Høgevold, Svensson, Padin, and Dos Santos [95] compare the difference between sustainable business models in manufacturing companies and hotels as a service sector. Their findings reveal that the nature of the industries is very effective in the models they choose to meet the sustainability objectives. Tangibility and intangibility of the products and services influence the ability of evaluation of the impact of their economic activities on the society and the environment.

Results of the research of Melissen, Cavagnaro, Damen, and Düweke [96] disclose that the current business models of hotel industry are not able to meet the sustainability objectives, especially with respect to addressing guests' needs and wants and (subsequent) institutionalization of sustainability. Nonetheless, they argue that managers' willingness and capabilities are potentially the sources stimulating them to transit toward a sustainable business model.

Høgevold and Svensson [97] develop a sustainable business model for the hotels based on a case study they have conducted among a major Scandinavian hotel chain known for having implemented sustainable business practices within the company and in its business network. They are the only study that has provided sustainable practical solutions for different elements of the business model in which the benefits of multi-stakeholders have considered in value creating and capturing processes.

**Table 15.** Application of Sustainable business models in the hospitality industry.

| Author/s | Years | Contribution | Methodology | Data Source |
|---|---|---|---|---|
| Buffa et al. [94] | 2018 | Empirical Evidence | Quantitative | Questionnaire |
| Høgevold et al. [95] | 2016 | Empirical Evidence | Qualitative | Case study, secondary data, company records, internet information, interviews, and on-site observations. |
| Melissen et al. [96] | 2016 | Empirical Evidence | Qualitative | Literature synthesis, Interviews |
| Høgevold and Svensson [97] | 2015 | Empirical Evidence | Qualitative | Case study, secondary data, company records, internet information, interviews, and on-site observations. |

**4. Discussion**

This study provides a comprehensive review of the applications of sustainable business models in different industries, sectors, and research area. Energy, fashion, healthcare, food, construction, and hospitality are industries have resorted to the principles of sustainable business models for the realization of sustainable development. Entrepreneurship, management and marketing, innovation, circular economy, and supply chain management are research areas that have utilized sustainable business models to provide solutions to achieve their sustainability ambitions. Application of sustainable business models in the developing countries is another category that has emerged in the initial screening phase of the literature.

Taking a closer look at the tables 2 to 15 reveals that circular business models, the base of the pyramid, and product service systems, are the major strategies that have been considered in the literature to design sustainable business models which are quite consistent with the findings of Bocken *et al.* [14]. Many authors consider designing a sustainable value proposition as an approach to design a sustainable business model. In this regard, Geissdoerfer *et al.* [26], Biloslavo *et al.* [27], Oskam *et al.* [28], Tolkamp *et al.* [34], Baldassarre *et al.* [35], De Bernardi and Tirabeni [36], and Hirscher *et al.* [51] have presented innovative approaches in which customers are engaged in the design process to devise a sustainable value proposition. Hirscher *et al.* [51], for instance, utilize do-it-yourself (DIY) and do-it-together [52] design strategies to design a more sustainable innovative value proposition. Geissdoerfer *et al.* [26], inspired by design thinking, developed the concept of value ideation to design a sustainable value proposition comprising additional benefits of stakeholders (including society and environment) in the value proposition. Oskam *et al.* [28] propose the concept value shaping to develop financial, social and environmental value that business creates by interacting with the different networks.

Studying the role of managers in designing sustainable business models is a topic has been considered in the common area of the literature of business and management and business model sustainability. Kurucz *et al.* [33] argue that relational leadership processes that support strongly sustainable organization management help organizations to meet their sustainability ambitions. On the other hand, Stubbs [31] believes that those organizations that have embedded the social and environmental concerns in their mission and their purpose have been successful in achieving their sustainable business model goals.

The main issue that has emerged in the application of sustainable business models in entrepreneurship is that despite there being a remarkable demand on the sustainable businesses, sustainable entrepreneurs are underrepresented [41] and the investors are reluctant to invest in e sustainable start-ups, particularly those that are environmentally sustainable [41]. Davies and Chambers [37] and Gasbarro *et al.* [38], on the other hand, consider business model sustainability innovation as the solution to conquering the barriers to implementing a sustainable business model.

Much research is conducted on developing solutions for sustainable business models to manage the resource and the energy as Moschetti *et al.* [42], Sousa-Zomer and Cauchick Miguel [43], and Zhang *et al.* [45] propose frameworks and approaches to develop sustainable business models to provide value to the energy resources. Rossignoli and Lionzo [46] also recommend partnership network is a solution that assists businesses in the energy sector to provide sustainable value propositions.

Pal and Gander [48] also believe that the traditional business models in the fashion industry produce highly negative outcomes for the environment through high water usage, chemical pollutions, and incineration or landfill of large amounts of unsold stock. Therefore, sustainable business models are considered as a solution to minimize such negative effects. The most prevalent approach in designing a sustainable business model in the fashion industry is the participation of the customers in the value creation process [51],[53]. On the other hand, Pal and Gander [48] believe that creating value for the customer and environment and capturing value for the firm is the solution to eliminate the barriers in the transition of traditional supply chain toward a slow approach and closing the loop of resources and a sustainable business model in the fashion industry.

Healthcare is another industry that utilizes sustainable business models to achieve sustainability goals. Surprisingly enough, the found articles have used sustainable business models to address sustainability issues have aided digital technologies. In other words, the common literature of business model sustainability and healthcare are tied with digital technology. Merchant *et al.* [55] and Anwar and Prasad [56] consider Telemedicine as a solution to design value propositions to develop a sustainable business model in the healthcare industry. In addition, Nikou and Bouwman [54] believe that utilizing mobile technology can design a sustainable business model in the healthcare industry.

The supply chain sector is the other sector that is found in the literature which has utilized the principles of the sustainable business model to provide solutions to enhance sustainable development [61].

Ray & Mondal [62], Geissdoerfer *et al.* [63], and Brennan and Tennant [64] argue that collaboration and networks among firms within a closed-loop supply chain (CLSC) leads to a sustainable business model in providing benefits to three bottom-line concepts to protect the environment, improve economic performance, and also social performance. Since the supply chain concept implies B2B relationships between the suppliers and buyers, such networks and collaboration can result in quarantinable consumption and according to Witjes and Lozano [61], it reduces the use of raw material and waste generation also. Finding exposes that the Bottom of the pyramid is the main approach for designing sustainable business models in developing countries. It is found that sustainable business models offer solutions such as designing a market-oriented business model to provide win-win solutions for multiple stakeholders. The research in the common field of business model sustainability and hotels, as the most important sectors of the hospitality industry, are still in the infancy stage as most of the studies have tried to investigate the sustainability level of the hotels rather than providing solutions for the development of a sustainable business model.

Illustrating the research path and articulating in detail the application of sustainable business models in different industries, sectors, and research area are the contributions of this study that provide insights and the possibility of compressions for both practitioners and researchers who are eager to find sustainable solutions through sustainable business models. Different approaches are proposed in the literature for designing a sustainable business model. Designing a sustainable value proposition which is able to provide values to multi-stakeholders such as society and environment while it can be profitable for the organization is the most common approach. Having a holistic view on the presented approaches reveals that designing a sustainable value creation, designing a sustainable value delivering, and sustainable B2B partnerships are other solutions have emerged in the literature for developing a sustainable business model.

## 5. Conclusions

The process of sustainable business model construction forms an innovative part of business strategy. Different industries and business types have utilized sustainability business models to satisfy their economic, environmental and social goals simultaneously. This study is conducted to present the state of the art of sustainable business models in various application areas. The business models are classified and reviewed in different application groups. To do so, a review is conducted, and the findings reveal that the application of sustainable business models can be classified in fourteen unique categories, which are innovation, management and marketing, entrepreneurship, energy, fashion, healthcare, agri-food, supply chain management, circular economy, developing countries, engineering, construction and real state, mobility and transportation, and hospitality industry. The study provides an insight into the state of the art of sustainable business models in various application areas and its research path. The main contribution of this study is the presentation of various applications of sustainable business models in different industries, sectors, and research areas. This study also provides insights for both practitioners and researchers to design a sustainable business model in different contexts. Many studies have proved and named the advantages sustainable business models have for organizations [8, 98] which might be led in a sustainable competitive advantage. On the other hand, there are external pressures and motivations from international organizations and NGOs on the organizations to be thrilled to shift toward sustainability. Therefore, the application of sustainable business models is increasingly widespread among different industries and sectors. Subsequently, dramatic advances in both research and practice have seen in the field of sustainable business models in different sectors. Hence, the reconnaissance of suitable strategies and innovation processes are initial actions for a transition toward a sustainable business model. The current study, bringing together the latest approaches which different sectors and industries take to transfer to a sustainable business model, provides managers with an insight into the advancements in this area as well as the possible solutions to facilitate the transition from a non-sustainable business model to a sustainable business model.

It is found that sustainable business models offer solutions such as designing a market-oriented business model to provide win-win solutions for multiple stakeholders. The research in the common field of business model sustainability and hotels, as the most important sectors of the hospitality industry, are still in the infancy stage as most of the studies have tried to investigate the sustainability level of the hotels rather than providing solutions for the development of a sustainable business model. The presented research in this article clarifies that four main approaches are emerged in the literature for designing a sustainable business model: designing a sustainable value proposition, designing a sustainable value creation, designing a sustainable value delivering, and finally generating sustainable partnership networks for creating and delivering such sustainable value which can meet the social, environmental and economic benefits at the same time.

An in-depth analysis of processes of transition from a traditional business model to a sustainable business model in different industries is recommended for future research. As can be seen in tables 2 to 15, most of the research has utilized a qualitative approach. Utilizing quantitative methodology to study the restrictive factors inhabiting the businesses to implement a sustainable business model and their effects on the social and environmental performance of the business is also recommended for future research.

**Conflicts of Interest:** Authors have no conflicts of interests.

Refrences